\documentstyle[12pt,epsfig]{article}
\begin{document}
\begin{center}

{{\large{ Resonant Diffraction Radiation
                 and Smith-Purcell Effect}}}
\end{center}
\vspace{5mm}
\begin{center}
{\large\sl  A.P.Potylitsin } \\
\end{center}
\begin{center}
{\small\it      Nuclear Physics Institute,\\
               Tomsk Polytechnic University, \\
               pr. Lenina 2A, P.O.Box 25, \\
               634050 Tomsk, Russia \\
               e-mail: pap@phtd.tpu.edu.ru
}
\end{center}

\begin{abstract}

     An approach has been developed where the
Smith-Purcell radiation  (SPR), i.e. emission of electrons
moving close to a periodic structure, is treated as
the resonant diffraction radiation. Simple formulas have been
designed for the SPR intensity for a grating having perfectly
 conducting strips spaced by a vacuum gap. The results have
 been compared with those obtained via other techniques.
 It has been shown that the intensity of radiation for the said gratings 
for a relativistic case sufficiently exceeds the SPR intensity for the 
grating made up by  a periodically deformed continuous surface.

\end{abstract}

\newpage

Radiation of electrons passing in the vicinity of a periodic
 structure made of some conducting material, the so-called 
Smith-Purcell effect (SPE), has been studied by both experimental and 
theoretical scholars for over 40 years now since its discovery.
Nonetheless, only the recent years saw the first experiments using
relativistic beams [1-3]. Design and realization of such experiments
followed by comparison with numerical calculations
(see e.g. Refs.4 and 5) is retarded due to the  absence of a
simple physical model capable of accounting for the
emission characteristics in terms of periodic structure parameters.
Within the current approach [6,7], based on the scalar diffraction theory,
the complexity of numerical calculations, resulting from a merely optical
problem of the grating reflectivity on the whole, "overshadows" the physical
 mechanism of SPE. We believe that the technique proposed by M.Moran
[8] is deemed more efficient. The author considers SPE for a stack of
 semi-infinite foils located normal to the beam trajectory using the radiation
 characteristics on each of the foil and taking into account interference
 from all the stack.

The present paper offers a more feasible model treating SPE as the resonant
diffraction radiation (RDR) of a relativistic particle on a set of 
strips of a finite width.
Fig.1 shows two different geometries of the strip arrangement:
parallel (case $a$) and perpendicular (case $b$).

Characteristics of the diffraction radiation (DR) of the \\ motion
of a charged particle close to a perfectly  conducting semi-plane were 
derived  in Ref.9 by solving the Maxwell equations via the  Wiener-Hopf 
method. Following the latter  formalism [9],  one may write the 
spectral-angular density of DR (neglecting the terms of the $\gamma^{-2}$ 
order,  $\gamma $ being the Lorentz factor) as:

$$
\frac{d^2W_{\mid \mid }}{d\Omega d\omega } = \frac {\alpha} {2\pi ^2}
\frac{ \exp (- \displaystyle \frac \omega {\omega _c}\sqrt{ 1 +
\gamma ^2\xi ^2\sin ^2\theta })}{ 1 - \cos \theta } \eqno(1)
$$
For the case a) and
$$
\frac{ d^2W_{\perp }}{d\Omega d\omega } = \frac{ \displaystyle \alpha}
{4\pi ^2}\frac{\exp \left( -\frac \omega {\omega _c}\sqrt{1 +
 \gamma ^2\xi ^2\sin ^2\theta }
\right) }{\sin ^2\theta }\frac{1 - \sin \theta +
 2\gamma ^2\xi ^2\sin ^2\theta }{1 + \gamma ^2\xi ^2\sin ^2\theta } \eqno(2)
$$
for the case b).

The  system of units used throughout the paper is
 $\displaystyle \hbar = m_e = c = 1$.

In the expressions (1) and (2) $\alpha = 1/137$ is the fine structure constant,
 $\displaystyle \omega_c = \frac{\gamma}{2h}$
 is the DR critical energy,  $h$  is the impact parameter, $\lambda$ is wavelength of DR,
 $\displaystyle \omega = \frac{2\pi}{\lambda },$
      $\theta $ and $\xi$ are the polar ($\theta >> \gamma ^{-1}$)
 and azimuthal photon outgoing angles ($\xi << 1  $ and is measured
from the plane containing the electron momentum and  normal to the strip
 edge).

The DR intensity is determined by the characteristic factor
(for $ \xi $ = 0)

$$ exp (- \frac{\omega}{\omega_c}) = exp (-\frac{4\pi h}{\gamma\lambda})$$

which can be connected with the intensity of virtual photons scattered by
the only edge of the semi-plane (the field strength of the virtual photons
 is determined by the factor
$\displaystyle exp (-\frac{2\pi h}{\gamma\lambda})).$

 It can be shown that, with electrons moving parallel to the strip surface,
 DR is formed during scattering of the virtual field on the edges
 of the strip, with
$$
\overrightarrow E_{exit} = - \overrightarrow E_{entrance} e^{i\varphi}
 \eqno (3)
$$

The interference of the fields (3) is taken into account in a manner similar to the
theory of transition radiation (TR) (see e.g. Ref.10):

$$
\frac{d^2W_{\mid \mid }^{st}}{d\omega d\Omega }  = \frac{d^2W_{\mid \mid }}{
d\omega d\Omega }\left| 1-e^{i\varphi _{\mid \mid }}\right| ^2 = \frac{
d^2W_{\mid \mid }}{d\omega d\Omega }F_{\mid \mid } \eqno(4)
$$
The interference factor

$$F_{\mid \mid } = 2(1- \cos\varphi_{\mid \mid }) =
4\sin^2\frac{\varphi_{\mid \mid }}{2}
\eqno(5)$$

is determined by the phase $\varphi_{\parallel}$ that coincides with the
 corresponding phase for TR of the vacuum slit and large outgoing angles
 ($\theta >> \gamma^{-1}$):

$$
\varphi _{\mid \mid } = \frac{2\pi a }\lambda \left( \frac 1\beta -\cos \theta
\right) \approx \frac{2\pi a }{\lambda} \left( 1-\cos \theta \right) \eqno(6)
$$

 In  the case $b$ we encounter a more complicated interference with fields of
 different strength (see Fig.1b):

$$
\frac{d^2W_{\perp }^{st}}{d\omega d\Omega } = \left| \overrightarrow{E}_{\perp
}\left( h-\frac a 2\right) e^{-i\varphi _{\perp }}-\overrightarrow{
E_{\perp }}\left( h+\frac a2\right) e^{+i\varphi _{\perp }}\right| ^2 =
$$
$$
= \left| \overrightarrow{E_{\perp }}(h)\right| ^2\left| \exp (\chi
- i\varphi _{\perp }) - \exp (- \chi +i\varphi _{\perp })^{}\right| ^2 =
  \frac{d^2W_{\perp }}{d\omega d\Omega }F_{\perp }, \eqno  (7)$$

$$
F_\bot = 4(\sin h^2\chi + \sin^2\varphi_\bot) \eqno(8)
$$

The notation used is as follows:

$$
\chi = \frac{\pi a}{\gamma \lambda }\sqrt{ 1 +\gamma^2\xi ^2\sin^2\theta}
\eqno (9)$$
$$
\varphi _{\perp }=\frac{\pi a}{\lambda} \sin \theta  \eqno (10) $$

It should be underlined that the interference factors (5) and (8)
are independent of the impact parameter $h$  as expected.

The interference of DR from a grating consisting of $N$ structural
elements (RDR) is described in an  entirely the same fashion as
is the resonant transition radiation [11]:

$$
\frac{d^2W_N^{\mid \mid ,\perp }}{d\Omega d\omega } = \frac{d^2W_{}^{\mid \mid
,\perp }}{d\Omega d\omega }F_{\mid \mid ,\perp }F_N,\eqno(11)
$$

$$
F_N = \frac{\sin ^2\left( N\frac{\varphi}2\right) }{\sin ^2\left( \frac{%
\varphi}2\right) }. \eqno (12) $$

The phase $\varphi$ in Eq. (12) is controlled solely by the grating period
$d$  but not its structure:
$$
\varphi  = \frac{2\pi d\left( \cos \theta - \frac 1\beta \right) }\lambda
\eqno(13)$$

For a considerably large number of elements, $ N  >>$ 1, we may use a simpler
 formula [10]:

$$F_N = 2\pi N\delta(\varphi-2k\pi), \eqno(14)$$
$k$ is the integer.

The argument of the $\delta$ - function governs the so-called
dispersion relation:

$$\lambda_k = \frac{d(\cos\theta - \frac{1}{\beta})}{k}  \eqno (15)$$

The above expression (15) is derived from general considerations irrespective
of any particular radiation mechanism.

   The expressions similar to Eq.(15) determine the position of lines in the
   spectrum radiated not only for SPE but also for RTR and parametric
  X-ray radiation [7,11,12].

Proceeding from Eqs.(4), (11) and (14), upon integration with respect to
frequency we may obtain the angular distribution of the SP radiation
calculated per  unit cell of the grating.

Consider first the case 1a). As follows from Eq.(5), the maximum yield
 is attained for the  $\varphi_{\mid \mid } = \pi$, where from  follows the
relation between the strip width and the period:

$$\frac{a}{d} = \frac{1}{2} \eqno (16)$$

Then, for the  fundamental diffraction order ($k$ = -1) we have:

$$
\frac{dW_{sp}^{\mid \mid _{}}}{d\Omega } = \frac{2\alpha }{\pi ^2}\omega _{-1}
\frac{\exp \left( - \frac{\displaystyle\omega _{-1}}{\omega _c}\sqrt{ 1 + \gamma ^2\xi ^2\sin
^2\theta }\right) }{ 1 - \cos \theta }, \eqno(17)
$$
where
$$
\omega _{-1} = \frac{2\pi }{\lambda _{-1}} = \frac{2\pi }{d\left( \frac 1\beta
- \cos \theta \right) }. \eqno(18)$$

In Ref.6 a similar problem was solved via  a different technique,
 the resulting Eq.(17), however, was rather close to the one in Ref.6,
  where instead of a two the numerator has the factor (1+$ \cos \theta$).

Fig.2a shows the calculations for the following conditions:
 $\gamma$ = 100; \  $d$ = 6mm; \  $h$ = 15mm;\  $\xi$ = 0.

For the 1b) geometry the SPR angular distribution under the same conditions
acquires a somewhat more complicated form:

$$
\frac{dW_{sp}^{\perp }(\xi=0)}{d\Omega } = \frac {\alpha}{\pi ^2}
\exp (-\frac{\omega_{-1}}{\omega _c})\omega _{-1}
\frac{1-\sin \theta }{\sin ^2\theta }(\sinh^2\chi +
\sin ^2\varphi _{\perp }) \eqno(19)
$$

The last term in Eq.(19) has a periodic character with the period being
 $\displaystyle \Delta\varphi_\bot = \pi$.
Therefore
the distribution Eq.(19) would have dips for the outgoing
angles, when  $\displaystyle sin^2\varphi_\perp $ becomes zero:

$$
\varphi _{\perp } = \frac{\pi a}{\lambda _{-1}}\sin \theta = \frac{\pi a\sin
\theta }{d( 1 - \cos \theta )} =  m\pi, \eqno (20) $$

Hence for  $\displaystyle \frac{a}{d} = \frac{1}{2}$ we have
 $\displaystyle \tan \frac{\theta_m}{2} = \frac{1}{2m}$,
and therefore,   $\displaystyle \theta_1= 53.13^{\circ}$;
$\displaystyle \theta_2=28.07^{\circ}$;
$\displaystyle \theta_3 = 18.92^{\circ}$; ... .

From  the relation Eq.(20) it also follows that the radiation intensity for a
constant outgoing angle will vary with $a/d$.
In Ref.6 this problem was analysed for the nonrelativistic case.

Using Eq.(2) and (11) we may now arrive at an azimuthal dependence
of the  SP  radiation yield for the angles 1$ >> \xi^2 >> \gamma^{-2}$:

$$
\frac{dW_{sp}^{\perp }}{d\Omega} \approx \frac {\alpha} {\pi ^2}\frac{2\omega _{-1}}{%
\sin^2 \theta } \exp \left( - 4 \frac {h}{a}\varphi _{\perp } \sqrt{\frac{\gamma ^{-2}%
}{\sin ^2\theta }+\xi ^2} \right) \times
\left( \varphi_{\perp }^2\xi ^2+\sin ^2\varphi_{\perp } \right)\;.
\eqno (21)
$$

It is obvious that in the case where
 $\displaystyle \sin^2\varphi_{\bot }  << \varphi^2_{\bot } $
(i.e. near $\varphi_{\bot } = m\pi$)
 the azimuthal dependence (21) would have the minimum for $\xi= 0$
 and  peak at  $\displaystyle \xi \sim  \frac{a}{2h\varphi_{\bot }} >>
 \gamma^{-1} $.
On the contrary, for parallel case distribution  (17) has only one 
maximum at $\xi$ = 0. Fig.3 presents the calculations for both geometries.
 In the relativistic case one can always choose the geometry and the 
grating parameters such that the radiation would be  concentrated near 
the plane normal to the grating.  As a result, first, the radiation 
brightness $ dW /d\Omega$ increases and, second, configuration of the
 would-be resonators and mirrors is simplified to an advantage as compared 
to that proposed in Ref.13, which may turn  out quite useful assuming the 
construction of an SPE - based free electron  laser.

The following should be noted in conclusion. All the experiments on SPE
studies of which the author is aware have been performed with volume gratings,
i.e. when a beam of virtual photons is reflected from a continuous surface
deformed via a certain periodic law. For instance, in the case of a lamellar
grating each unit cell of the latter contains two perpendicular and two parallel
planes. Due to destructive interference the reflectivity $F$ of the grating
element ( $\vert R_{-1}\vert^2$ in terms of Ref.4) is sufficiently below 
unity ($10^{-2} \div 10^{-3}$, while for a grating of parallel strips 
separated by vacuum gaps, this quantity  may reach a value of 4 
(see Eq.(5)) and, therefore, the SPR intensity would grow in the  same fashion 
(increasing by 2-3 orders).

In the relativistic case the thickness is chosen such that
 $\gamma\lambda >> b$  at least in the infrared and millimeter range, 
where polished  aluminum could be considered as an ideal conductor.

Thus, for our case the model proposed is going to be sufficiently valid.
The deduction made by the authors of Ref.4 on the absence of advantages
for the relativistic beams to be used as the SPR generators seems to us
premature.

\vspace{5mm}
\hspace{8mm} The author greatly appreciates the contribution by \\
Prof.M.Ikezawa and Dr.Y.Shibata rendered through numerous fruitful 
discussions and stimulating criticism as well as the possibility of 
getting acquainted with their experimental results before  publication.

\newpage

\centerline{\large References}
\begin{enumerate}
\item
G.Doucas, J.H.Mulvey, M.Omori et al.  Phys.Rev.Lett., \underline{69},
\\ 1761(1992)
 \item K.J.Woods, J.E.Walsh, R.E.Stoner et al.  Phys.Rev.E., \underline{74},
 \\ 3808(1995)
  \item
  K.Ishi, Y.Shibata, T.Takahashi et al. Phys.Rev.E, \underline{51},
 \\ R5212(1995)
 \item
 O.Haeberle, P.Rullhusen, J.M.Salome et al.Phys.Rev.E,\underline{49},
 \\ 3340(1994)
 \item
 J.Walsh, K.Woods, S.Yeager. NIM A341, 277(1994)
 \item
 B.M.Bolotovski, G.V.Voskresenskii. Sov. Phys. Uspekhi,\underline{11},
 \\ 143(1968)
  \item
  M.L.Ter-Mikaelian. High-Energy Electromagnetic Processes in
  Condensed Media (Wiley Interscience, New York, 1972)
\item
M.J.Moran.  Phys.Rev.Lett. \underline{69}, 2523(1992)
\item
A.P.Kazantsev, G.I.Surdutovich. Sov.Phys.-Doklady,\underline{7}, \\ 
990(1963)
\item
X.Artru, G.B.Yodh, G.Mennessier . Phys.Rev.D,\underline{12}, \\ 
1289(1975)
\item
G.M.Garibyan, L.A.Gevorgyan, C.Yang. Sov.Phys.JETP,\underline{39},
\\ 265(1974)
\item
Yu.N.Adischev, V.A.Verzilov, A.P.Potylitsin et al.
NIM \underline{B44},\\ 130(1989)
\item
A.Gover, P.Dvorkis, U.Elisha. J.Opt.Soc.Am. B1, 723(1984)
\end{enumerate}

\newpage

\begin{center}
{\Large \bf                       Figure Captions}
\end{center}

Fig.1   Grating geometry with finite width strips : \\
        a) - parallel and \\
        b) - perpendicular configuration 

\vspace{3mm}
Fig.2   Angular distribution of SPR intensity : \\
        upper curve - geometry a), \\
        lower curve - geometry b). 

\vspace{3mm}
Fig.3   Azimuthal dependence of SPR intensity for polar angle \\
         $\theta=70^{\circ} $:

        upper curve - geometry a),

        lower curve - geometry b).

\newpage

\vspace{2cm}
\unitlength=1cm
\begin{picture}(18,30)
\put(2,13.5){\epsfxsize=10cm\epsfbox{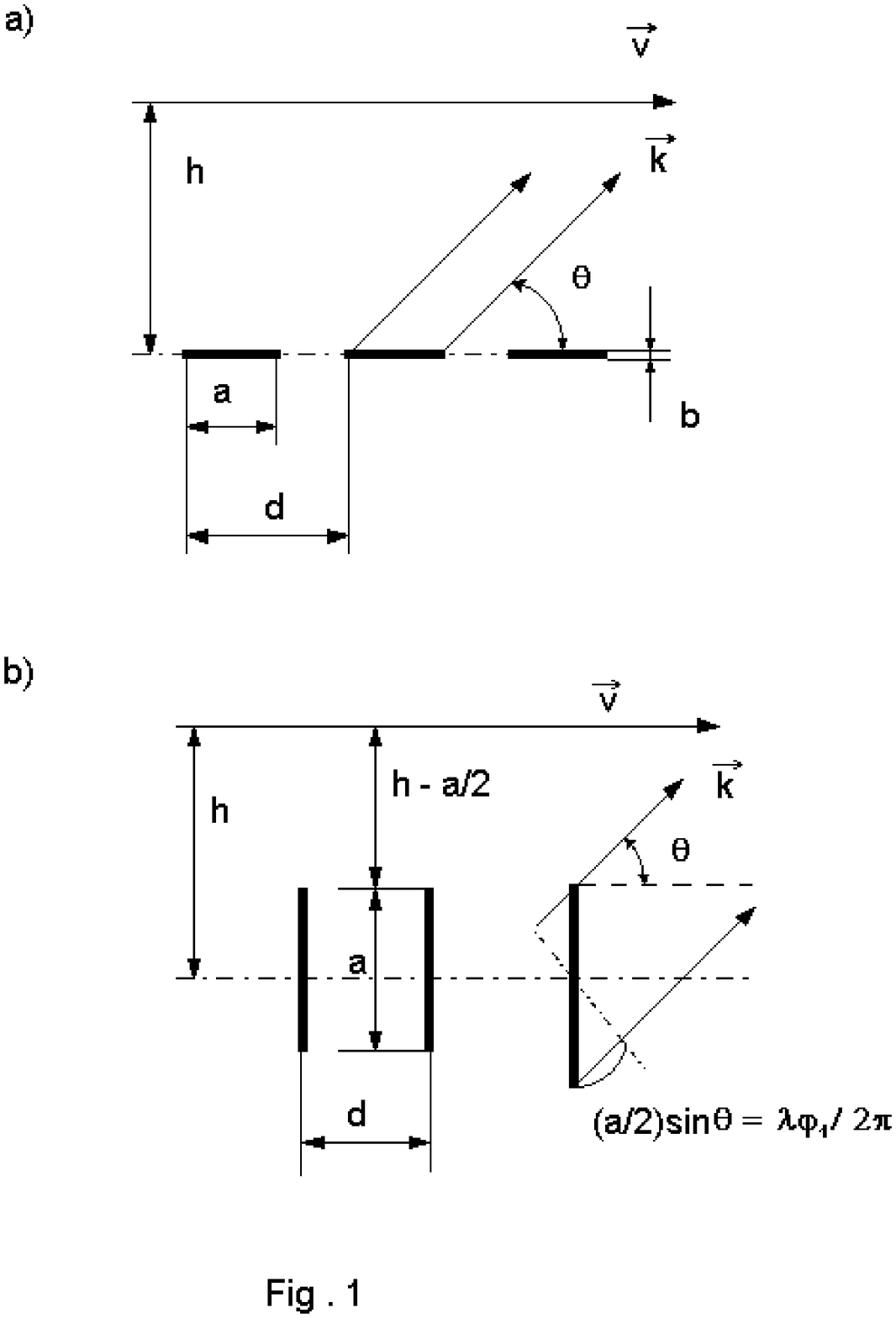}}
\end{picture}

\newpage
\vspace{2cm}
\unitlength=1cm
\begin{picture}(18,30)
\put(2,13.5){\epsfxsize=10cm\epsfbox{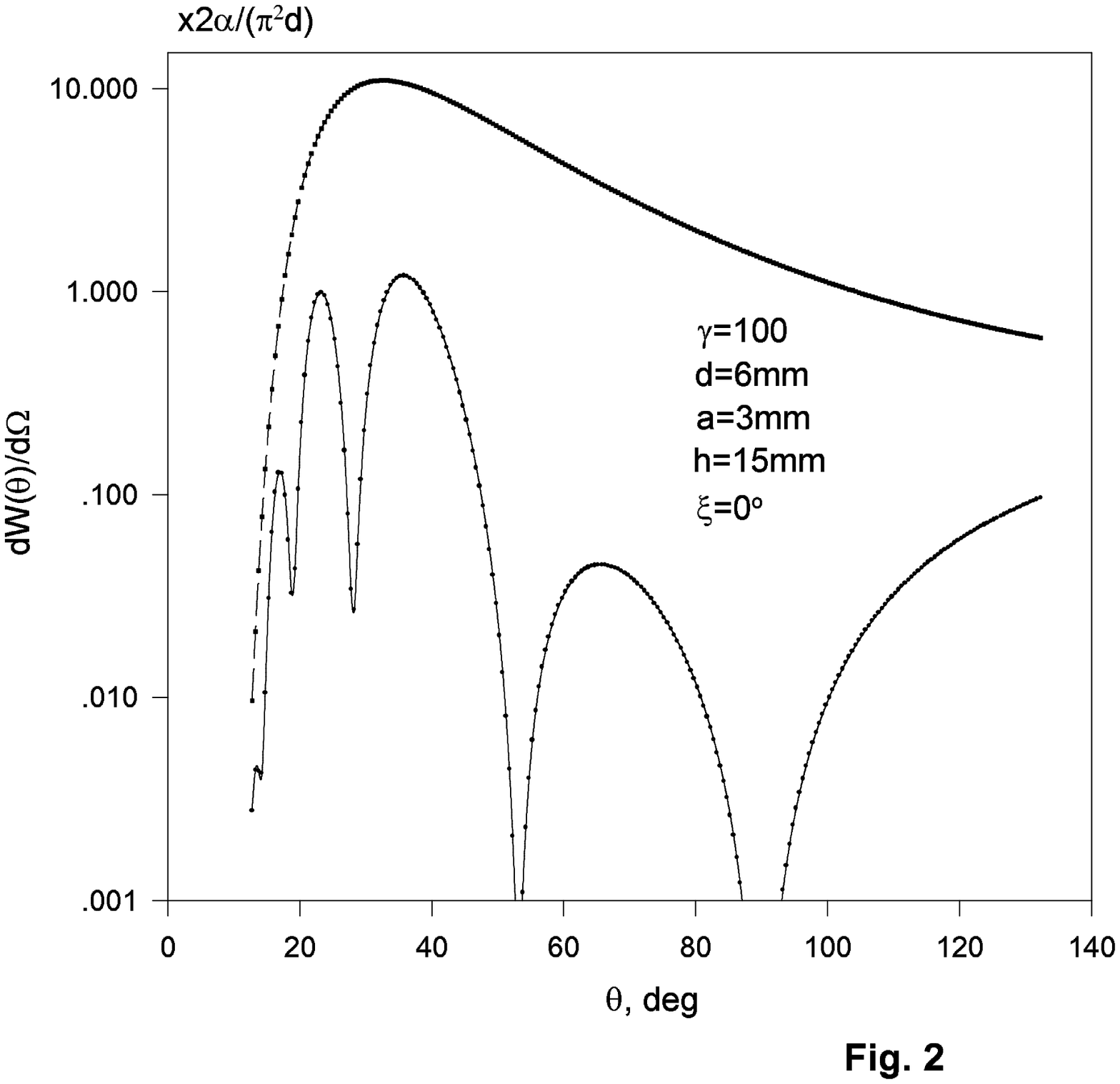}}
\end{picture}

\newpage
\vspace{2cm}
\unitlength=1cm
\begin{picture}(18,30)
\put(2,13.5){\epsfxsize=10cm\epsfbox{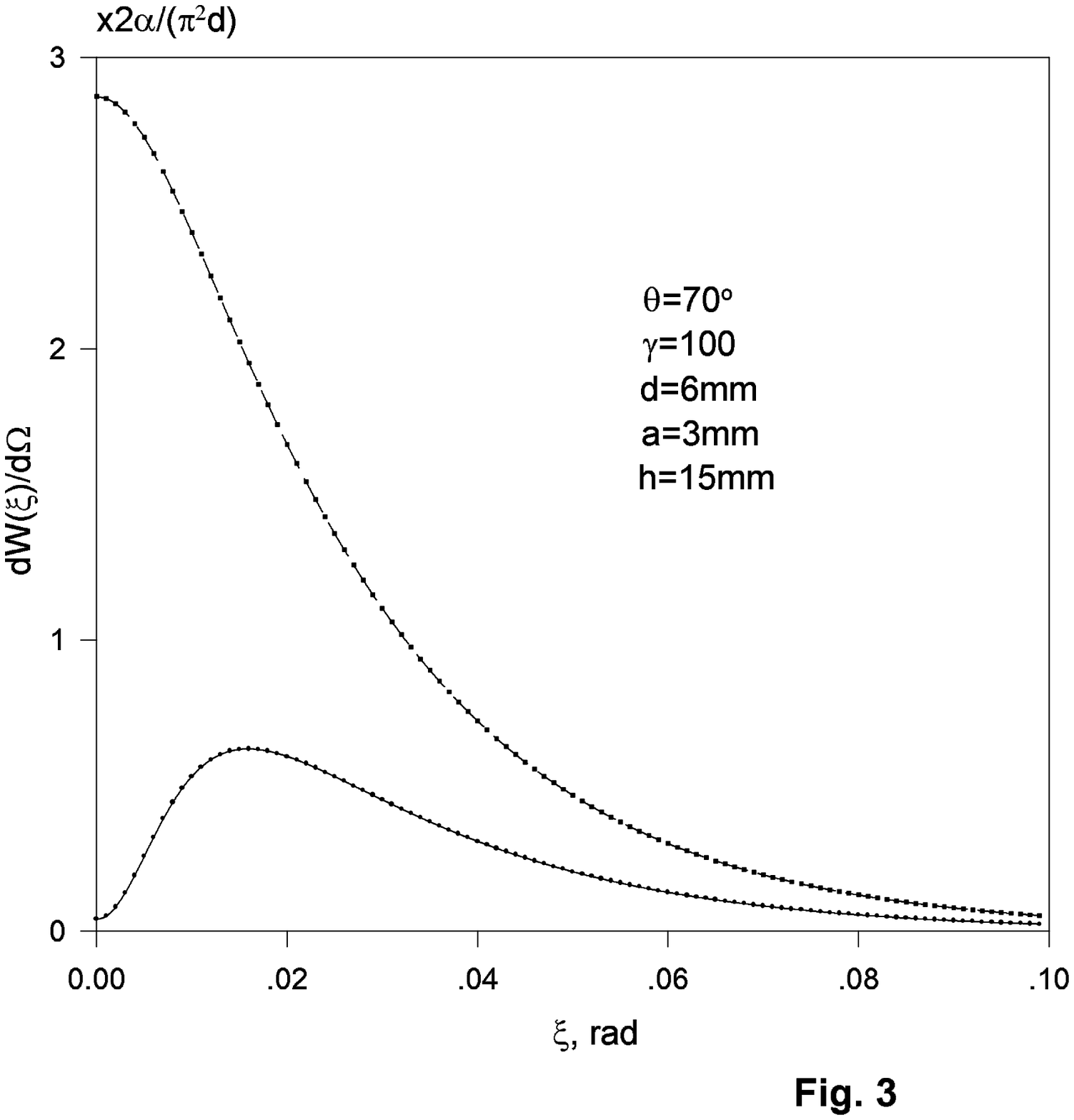}}
\end{picture}

\end{document}